\documentstyle[epsfig,graphicx,12pt,epsf]{article}

\begin{document}

\begin{center}
{\Large{\bf Deeply bound pionic atoms from the $(\gamma,p)$ reaction in nuclei
  }}

\vspace{0.3cm}

\end{center}

\vspace{1cm}

\begin{center}
{\large{S. Hirenzaki$^{1,2}$ and E. Oset$^{1}$}}
\end{center}

\begin{center}
{\small{$^{1}$ \it Departamento de F\'{\i}sica Te\'orica and IFIC, \\
Centro Mixto Universidad de Valencia-CSIC, \\
Ap. Correos 22085, E-46071 Valencia, Spain}}

{\small{$^{2}$ \it Department of Physics, Nara Women's University, Nara
630-8506, Japan}}

\end{center}

\vspace{1cm}

\begin{abstract}
  We study the $(\gamma,p)$ reaction on $^{208}Pb$ leading to $^{207}Pb$
  with a bound pion attached to it in the lowest 1s or 2p pionic levels.  The
  reaction can be made recoilless to optimize the production cross
  section but we must choose a bit higher photon energy to overcome the
  Coulomb barrier in the proton emission.  The cross sections obtained are
easily measurable
   and can be  larger than 50 per cent of the background from inclusive
   ($\gamma$,p). This makes it a clear
  case for the detection of the pionic atom signals, converting this
reaction into a
  practical tool to produce deeply bound pionic atoms.

\end{abstract}

\section{Introduction}
In the past decade a search, both theoretical and experimental, for deeply
bound states was conducted, which lead to the successful detection of these
 states in $Pb$ isotopes in \cite{exp1,exp2}. A review of the methods proposed
 and early attempts prior to the detection at GSI can be found in
\cite{hakone}.
 Out of many reactions proposed it was envisaged in \cite{hakone} that two
 reactions stood better chances, the $(d,^3He)$ reaction \cite{toki} and the
 radiative capture in low energy pion scattering $(\pi^-, \gamma)$ \cite{juan},
 both of them leaving a
 bound pion in the nucleus.  The first reaction offers a special
 characteristic, very dear in production of bound particles in nuclei, which is
 its recoilless nature for some kinematics. The second reaction is also nearly
 recoilless and has the additional advantage of being a coherent reaction,
hence
 benefiting from an extra $A^2$ factor in the cross section. This extra
 advantage is however counterbalanced by the the fact that the pions are
 secondary beams and hence one has smaller fluxes than with primary beams like
 the $d$ in the first reaction. The first reaction is the one that led to the
 successful detection of the pionic states, while the second one, carried
out at
 TRIUMF \cite{exp3}, offered a much less clear evidence because lack of enough
 resolution in the photon detector. Yet, it proved a worth method to be
 considered in the future when better resolution is achieved since it produces
 the pionic states in the ground state of stable nuclei, like $^{208}Pb$.
 The  $(d,^3He)$ reaction leads instead both to ground and excited states of
A-odd nuclei where the
 atomic and nuclear levels mix.

    The experience on the history of these reactions also stresses the
importance
 of having a recoilless reaction and using primary beams. Another reaction
which
 fulfills these conditions is the $(\gamma,p)$ reaction where a particular
 kinematics can be chosen to make the reaction recoilless in the production of
 the pionic atom.  A suggestion to use this reaction for the production of
omega
 bound states in nuclei has been done in \cite{uge}, although the large width
 predicted for these states would most probably prevent the observation of
clean
 peaks. In the present case we already know from experiment that the widths are
 smaller than the separation of the levels. The success of the reaction is then
 tied to the magnitude of the cross sections for the reaction and the ratio
 of signal to the
 background coming from other reactions where no pions are produced but a
 proton is detected. Indeed,
 consideration of the background is important since
 the signal for the bound pion production is the detection of a proton with an
 energy equal to the one of the photon minus the pion mass ( and the binding
 energies). However, the same protons can be obtained from the inclusive
$(\gamma,p)$
 process in which the proton has collisions and loses energy.

   In the present work we perform calculations for the reaction

\begin{displaymath}
	\gamma + ^{208}Pb \rightarrow p + ( ^{207}Pb \cdot \pi^-_b)
\end{displaymath}

\noindent
   which is depicted diagrammatically in fig. 1.  The elementary reaction is
   $\gamma n\to \pi^- p$ and the final proton is emitted leaving a nuclear
state
   of $^{207}Pb$ with a bound pion tied to it.

\begin{figure}
\begin{center}
\leavevmode
\epsfysize=7cm
\epsffile{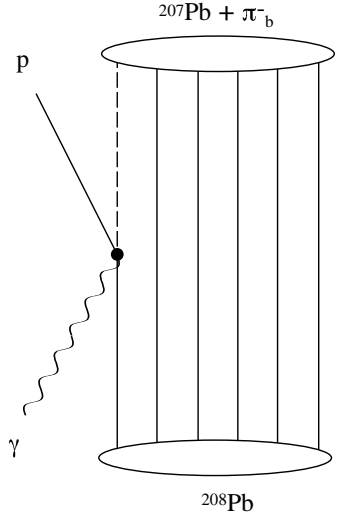}
\caption{Diagram for the ($\gamma$,p) reactions to form pionic bound
states on $^{207}$Pb. }
\label{fig:1}
\end{center}
\end{figure}

     Since the reaction is made practically recoilless, the elementary
amplitude for the
 process is extremely simple since only the Kroll Ruderman term contributes.
Thus
 the nuclear amplitude for the process of fig. 1 is given by

\begin{equation}
- i T = \int d^3r \frac{f_{\pi NN}}{m_{\pi}} e \sqrt{2} \left(
\frac{2M_N}{2M_N-m_{\pi}} \right) \vec{\sigma}\cdot \vec{\epsilon}
\frac{1}{\sqrt{V}} e ^{i\vec{k} \cdot \vec{r}}
\frac{1}{\sqrt{V}} e ^{-i\vec{p} \cdot \vec{r}}
\Phi ^*_{nlm} (\vec{r}) \phi_{JLM}  (\vec{r})
\end{equation}

\noindent
 where $\Phi$ and $\phi$ are the pion and neutron wave function
 respectively, $k$, $p$ the photon and proton momenta,
 and $\vec{\epsilon}$ the photon polarization.

   In eq. (1) we have used a plane wave for the proton, but it is clear that a
 distorted wave should be used and we shall do that below.
 In addition we
sum and average the cross section over final and initial polarizations.
We show explicitly below the results obtained with the photon
polarization given by
 $\vec{\epsilon}_-=(\hat{i}-i\hat{j})/\sqrt{2}$.

   We shall look at protons going in the forward direction.
The matrix element of $T$, removing the volume V in the
   denominator, becomes $T^{\prime}$ given by
\begin{eqnarray}
T^{\prime}  &=& 2 \pi \delta_{M-1/2, m} \frac{f_{\pi NN}}{m_{\pi}} 2 e
\frac{2M_N}{2M_N-m_{\pi}} \nonumber\\
&\times& \sum_{l_p} \int^{\infty}_0 b db \int^{\infty}_{-\infty}
dz \tilde{R}_{nl}^*(r) R_{NL}(r) \tilde{Y}_{l,m} (\frac{z}{r})
\tilde{Y}_{L,M-1/2} (\frac{z}{r}) \nonumber\\
&\times& C (L, 1/2, J; M-1/2,1/2) e^{ikz} (2l_p+1)(-i)^{l_p}
\tilde{j_{l_p}} (pr) P_{l_p} (\frac{z}{r})
\end{eqnarray}

\noindent
   where $\tilde{R}$ and $R$ are the radial wave functions of the pion and
   bound neutron states, and $\tilde{Y}$ the
   spherical harmonics removing the $e^{im\varphi}$ factor.
   The proton distorted wave $\tilde{j_{l_p}} (pr)$ is obtained for
   each proton partial wave by solving the Schr\"odinger equation with
   the appropriate
   boundary condition shown in Eq. (12) in Ref. \cite{nieves93}.
 We take the proton-nucleus optical potential from \cite{mahaux85}
 and it is given by

\begin{equation}
U(r) = \frac{V+iW}{1+exp[ (r-R)/a]}   ,
\end{equation}

\noindent
where $R$ and $a$ are the radius and diffuseness parameters taken to be
7.35 fm and 0.65 fm for Pb, respectively. The energy dependent potential
strength $V$ and $W$ are
   shown in Fig. 2.10 (for the real part)
   and Fig. 4.6 (for the imaginary part) in
   Ref. \cite{mahaux85}.
  We also add the Coulomb potential with finite size of the
   nucleus, which is the same one used for the $\pi^-$ with opposite sign.

     The cross section for the process is then given by

\begin{eqnarray}
\frac{d^2\sigma}{d\Omega dE_p} & = & \frac{pM_N}{2m_{\pi}}
\frac{1}{(2\pi)^3}
\frac{1}{2k} \nonumber\\
& \times &
\sum_{(n^{-1} \otimes \pi)}
\frac{\Gamma}{[k+M(^{208}Pb) - E_p - M(^{207}Pb\cdot \pi^-_{b})]^2 +
(\Gamma/2)^2}
\nonumber\\
& \times &
\bar{\sum} \sum | T^{\prime} |^2
\end{eqnarray}

\noindent
    where  $\Gamma$ is the width
     of the pionic atom state.

     By numerical calculations,
     we find that the effect of the Coulomb barrier for
     the emitted proton is large for low energy protons, suppressing the
     signals significantly. On the other hand, if we increase the proton
     kinetic energy in the final states, which is equivalent to increasing
     the incident photon energy, this moves us away from the recoilless
condition of the
     reaction.  Thus, we evaluate the pionic atom formation spectra for
several incident
     photon energies in order to find the optimal kinematical condition.
     We conclude that we have the largest signals at an incident photon
     energy $k=170MeV$, while the recoilless condition appears at
$k=155MeV$.
          The results for the double differential cross section of the
reaction are
     shown in fig. 2.
As we expect for the nearly recoilless kinematics, the substitutional states
make the largest peaks and the dominant contributions come from the 2p
pionic state with neutron p$_{3/2}$ and p$_{1/2}$ hole states.
We can
also see the pionic 1s state formation at T$_p$=29 $\sim$ 30 MeV.
The formation cross section is reduced by
around the 30 \% from the plane wave approximation for the largest signal
due to the distortion
effects on the emitted proton
and is estimated to be about 20 $\mu b/sr/MeV$ for the largest signal.


\begin{figure}
\begin{center}
\leavevmode
\epsfysize=12cm
\epsffile{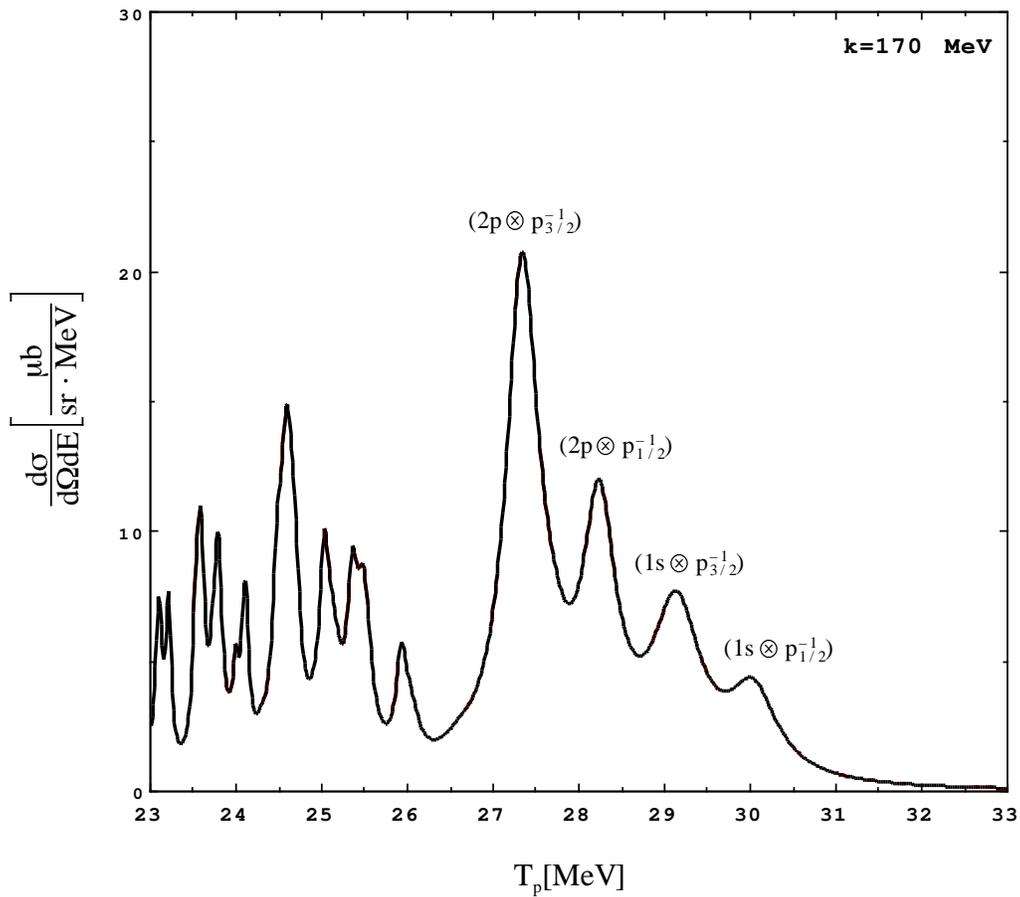}
\caption{Expected spectra of the $^{208}$Pb($\gamma$,p)$^{207}$Pb$\cdot
\pi^-_b$ reaction at the photon energy k=170 MeV as a function of the
emitted proton energy.  A convolution
with the experimental
resolution of 50 keV FWHM is implemented in the results.
For a resolution of 200 keV the figure is similar, but the strength at
the peak of the p-wave states is reduced by about 20 \%. }
\label{fig:2}
\end{center}
\end{figure}

  As for the background for the present reaction it can be easily estimated
using
  experimental results already available from the study of the inclusive
$(\gamma,p)$
  reaction in nuclei \cite{arends}. There we can see results in
  $^{208}$Pb at $k=227MeV$ and $390MeV$.  The cross sections go down to
  E$_p$=40MeV and hence it is easy to extrapolate the results smoothly to
  E$_p$=30MeV that we have in the present case. The differential cross
  sections for a proton angle of 52$^o$ are 50 $\mu b /sr/MeV$ and 100
  $\mu b /sr/MeV$ for $k=227MeV$ and $k=390MeV$, respectively.
  Extrapolating from these two data to $k=170MeV$ we obtain a cross
  section of 32 $\mu b /sr/MeV$ for the background of inclusive
  ($\gamma$,p) in our reaction at this angle.  This result could be also
  obtained theoretically using the approach of \cite{rafa} where
     a Monte Carlo simulation was performed with the probabilities for the
primary nuclear
  steps evaluated with a microscopic many body calculation in \cite{photon}.
   This inclusive ($\gamma$,p) cross section is about double the one obtained
  for the peaks of the signals of the pionic atoms in an average.
   However, at small angles,
  where we have evaluated the pionic atom formation cross sections, the
background should be
  even smaller, and this is the case in the theoretical model of \cite{rafa},
  because kinematically it is not possible to have
  contribution to the process with just one collision.

  This situation is
  particularly rewarding in view to distinguish these states experimentally.
  On the other hand, although our calculations do not deem it necessary,
  it should
  be possible to further increase the ratio of signal to background using
delayed
  fission of pion absorption fragments, a technique used with success in
  the production of $\Lambda$ hypernuclei in
  \cite{hyp} and tentatively suggested for the present reaction \cite{amour}.
  The set up for the experiment is suited to present experimental
  facilities, particularly those of low energies, and the cross sections
  predicted are of the order of those presently being measured with high
  precision in the ($\gamma$,p) reaction \cite{arends}, which makes
  this reaction a very practical method to produce deeply bound pionic
  atoms.

\subsection*{Acknowledgments}

One of us, S.H. wishes to acknowledge the hospitality of the University
of Valencia where this work was done and financial support from the Fundacion
BBV.
We would like to thank A. Margarian for useful discussions and
encouragement to perform these calculations.
This work is also
partly supported by DGICYT contract number BFM2000-1326 and
by the Grants-in-Aid for Scientific Research of the Japan Ministry of
Education, Culture, Sports, Science and  Technology (No. 11440073 and No.
11694082).

\end{document}